\documentclass[submission,copyright,creativecommons]{eptcs}
\providecommand{\event}{GandALF 2022} % Name of the event you are submitting to
\usepackage{underscore}
\usepackage{amssymb} %, amsthm}
\usepackage{url}
\usepackage{amsmath}
\usepackage{cite}
\usepackage{cleveref}
\usepackage{latexsym, graphics, graphicx}
\usepackage{color}
\usepackage{epsfig, amssymb}
\usepackage{yfonts}
\usepackage{algorithmic}
\usepackage{algorithm}
\usepackage{float}
\usepackage{enumerate}
\usepackage{hyperref}
\usepackage{mathtools}
\usepackage{todonotes}
\usepackage{caption}
\usepackage{subcaption}
\usepackage{listings}
\usepackage{tikz}
\usepackage{siunitx}
\usepackage{pgfplots}
\usetikzlibrary{arrows, backgrounds, fit, decorations.pathreplacing, positioning, shadows}

\def\orcidID#1{\unskip$^{[#1]}$}

\usepgfplotslibrary{units}

\definecolor{keyblue}{rgb}{0.1, 0.1, 0.6}
\definecolor{dkgreen}{rgb}{0,0.6,0}
\definecolor{gray}{rgb}{0.5,0.5,0.5}
\definecolor{stringcol}{rgb}{0.58,0.4,0.1}

\DeclareCaptionFont{white}{\color{white}}
\DeclareCaptionFormat{listing}{\colorbox[cmyk]{0.73, 0.35, 0.15,.5}{\parbox{\textwidth}{#1#2#3}}}
\newtheorem{example}{Example}

\newcommand{\ignore}[1]{}

\newcommand{\E}{\mathcal{E}}
\newcommand{\D}{\mathcal{D}}

\title{CryptoSolve: Towards a Tool for the Symbolic Analysis of Cryptographic Algorithms}

\author{
	Dalton Chichester
	\institute{University of Mary Washington, Fredericksburg, VA, USA} 
	\email{dchiches@mail.umw.edu} \and
	Wei Du
	\institute{University at Albany--SUNY, Albany, NY, USA}
	\orcidID{0000-0002-9149-6229} 
	\email{wdu2@albany.edu} \and
	Raymond Kauffman
	\institute{University of Mary Washington, Fredericksburg, VA, USA} 
	\email{rkauffma@mail.umw.edu} \and
	Hai Lin
	\institute{Clarkson University, Potsdam, NY, USA}
	\orcidID{0000-0001-8658-9634} 
	\email{hlin@clarkson.edu} \and
	Christopher Lynch
	\institute{Clarkson University, Potsdam, NY, USA}
	\orcidID{0000-0003-1141-0665} 
	\email{clynch@clarkson.edu} \and
	Andrew M. Marshall
	\institute{University of Mary Washington, Fredericksburg, VA, USA}
	\orcidID{0000-0002-0522-8384} 
	\email{amarsha2@umw.edu} \and
	Catherine A. Meadows
	\institute{Naval Research Laboratory, Washington, DC, USA} 
	\email{catherine.meadows@nrl.navy.mil} \and
	Paliath Narendran
	\institute{University at Albany--SUNY, Albany, NY, USA}
	\orcidID{0000-0003-4521-5892}  
	\email{pnarendran@albany.edu} \and
	Veena Ravishankar
	\institute{University of Mary Washington, Fredericksburg, VA, USA}
	\orcidID{0000-0003-3498-4039} 
	\email{vravisha@umw.edu} \and
	Luis Rovira
	\institute{University of Mary Washington, Fredericksburg, VA, USA}
	\email{lrovira@umw.edu} \and
	Brandon Rozek
	\institute{Rensselaer Polytechnic Institute, Troy NY, USA}
	\orcidID{0000-0002-4537-559X}
	\email{rozekb@rpi.edu}
}

\newcommand{\titlerunning}{CryptoSolve}
\newcommand{\authorrunning}{Chichester, et al.}

\begin{document}
\maketitle

\begin{abstract}
	Recently, interest has been emerging in the application of symbolic techniques
	to the specification and analysis of cryptosystems.
	These techniques, when accompanied by suitable proofs of soundness/completeness,
	can be used both to identify insecure cryptosystems and prove sound ones secure.
	But although a number of such symbolic algorithms have been developed and implemented,
	they remain scattered throughout the literature.
	In this paper, we present a tool, CryptoSolve, which provides a common basis for specification
	and implementation of these algorithms, CryptoSolve includes libraries that provide the
	term algebras used to express symbolic cryptographic systems,
	as well as implementations of useful algorithms, such as unification and variant generation.
	In its current initial iteration, it features several algorithms for the generation
	and analysis of cryptographic modes of operation, which allow one to use block ciphers to encrypt messages more than one block long.
	The goal of our work is to continue expanding the tool in order to consider additional cryptosystems and security questions,
	as well as extend the symbolic libraries to increase their applicability.
\end{abstract}

\section{Introduction}

Although security properties of cryptographic algorithms are generally proved using a computational
model in which probabilities of events are explicitly quantified, there are often advantages to using a
more easily automated abstract symbolic model.
This is particularly the case when one is looking for cryptosystems that obey some non-cryptographic constraints,
e.g. parallelizability, or even non-technical constraints, such as absence of intellectual property restrictions.
One can use automated both methods to generate a large number of candidate cryptosystems,
and to verify the security in the symbolic model.  If the symbolic model is computationally sound
(that is if the symbolic analogue of a particular security property holds) it is possible to use this
technique to identify secure cryptosystems.
Even the symbolic model is not computationally sound, but is computationally complete, it is possible to use
the technique to weed out insecure constructions.
A growing body of work, e.g.\cite{DBLP:conf/ccs/BartheCGKLSB13,DBLP:conf/csfw/MalozemoffKG14,10.1145/2810103.2813636,DBLP:conf/crypto/CarmerR16}],
shows how this can be applied to the construction of new cryptosystems.
Symbolic methods can also be useful by themselves,  even without automatic generation.
For example,  in \cite{DBLP:conf/ctrsa/VenemaA21} Venema and Alp{\'{a}r}
use symbolic methods to find security flaws in recently proposed attribute-based encryption schemes,
in \cite{Hollenberg22}  Hollengberg, Rosulek, and  Roy  and in \cite{MeadowsESORICS21} Meadows respectively
find different symbolic criteria guaranteeing the security of blockcipher cryptographic modes of operation
under different usage assunmptions, and in \cite{DBLP:conf/tcc/McQuoidSR19} McQuoid, Swope, and Rosulek develop
a polynomial-time algorithm for checking security properties of a class of hash functions.

However, the
symbolic problems we encounter often come with constraints tied to the properties of the cryptosystem,
such as, requiring that any substitutions be constructible from terms and function symbols available to
an adversary, or that the adversary may not be able to perform certain operations,
such encryption with a key that is not available to it. Hence specialized algorithms or tools may be necessary.

In this paper we present an overview of an initial version of such a tool,
CryptoSolve,\footnote{The current version of the tool can be found here: \url{https://symcollab.github.io/CryptoSolve/}.}
that has been designed to generate and analyze specifications of cryptosystems.
This in turn allows for the automatic generation and symbolic analysis of certain cryptographic algorithms.
The goal of this new tool is broad, to develop not only a usable analysis tool for an extensive family of cryptographic algorithms
but to also develop the underlying libraries which could be used in analysis of
additional algorithms, properties, and within other symbolic analysis tools.

Our initial version of CryptoSolve provides algorithms for reasoning about the security and functionality
of a class of cryptosystems known as \emph{cryptographic modes of operation}.
These modes use fixed length block ciphers to encrypt messages more than one block long.
The key property of a mode is to protect the secrecy of the encrypted data,
but it may also be used to provide integrity and authentication.
In addition, the mode must be invertible by anyone who possesses the decryption key.
CryptoSolve supports both the automated and manual generation of modes.
It also features algorithms for checking secrecy
(in the sense of indistinguishability of ciphertext from random), authenticity, and invertibility.

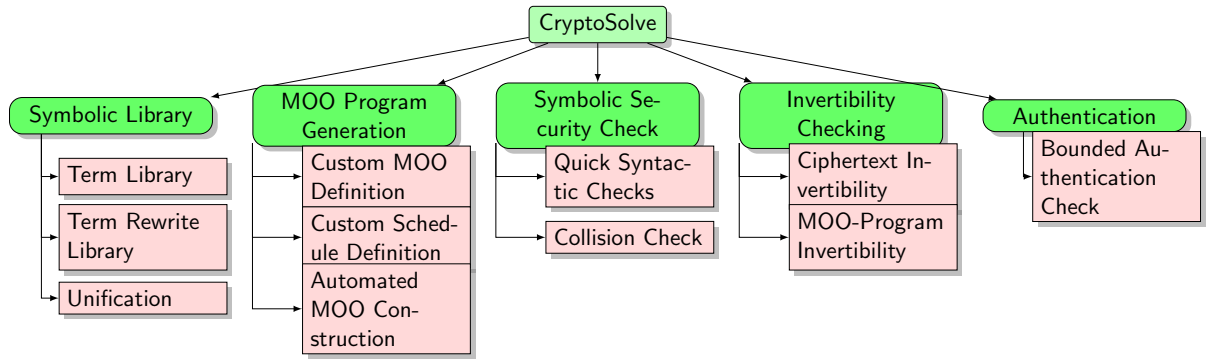
\begin{figure}
	\resizebox{\textwidth}{!}{
	\centering
	\begin{tikzpicture}[
	level 1/.style={sibling distance=40mm},
	edge from parent/.style={->,draw},
	>=latex]

	\tikzset{
		basic/.style  = {draw, text width=2cm, drop shadow, font=\sffamily, rectangle},
		root/.style   = {basic, rounded corners=2pt, thin, align=center,
			fill=green!30},
		level 2/.style = {basic, rounded corners=6pt, thin,align=center, fill=green!60,
			text width=8em},
		level 3/.style = {basic, thin, align=left, fill=pink!60, text width=6.5em}
	}

	% root of the the initial tree, level 1
	\node[root] {CryptoSolve}
	% The first level, as children of the initial tree
	child {node[level 2] (c1) {Symbolic Library}}
	child {node[level 2] (c2) {MOO Program Generation}}
	child {node[level 2] (c3) {Symbolic Security Check}}
	child {node[level 2] (c4) {Invertibility Checking}}
	child {node[level 2] (c5) {Authentication}};

	% The second level, relatively positioned nodes
	\begin{scope}[every node/.style={level 3}]
	\node [below of = c1, xshift=15pt] (c11) {Term Library};
	\node [below of = c11] (c12) {Term Rewrite Library};
	\node [below of = c12] (c13) {Unification};

	\node [below of = c2, xshift=15pt] (c21) {Custom MOO Definition};
	\node [below of = c21] (c22) {Custom Schedule Definition};
	\node [below of = c22, yshift=-5pt] (c23) {Automated MOO Construction};

	\node [below of = c3, xshift=15pt] (c31) {Quick Syntactic Checks};
	\node [below of = c31] (c32) {Collision Check};

    \node [below of = c4, xshift=15pt] (c41) {Ciphertext Invertibility};
    \node [below of = c41] (c42) {MOO-Program Invertibility};
    
    \node [below of = c5, xshift=15pt] (c51) {Bounded Authentication Check};

	\end{scope}

	% lines from each level 1 node to every one of its "children"
	\foreach \value in {1,2,3}
	\draw[->] (c1.195) |- (c1\value.west);

	\foreach \value in {1,2,3}
	\draw[->] (c2.195) |- (c2\value.west);

	\foreach \value in {1,2}
	\draw[->] (c3.195) |- (c3\value.west);

	\foreach \value in {1,2}
	\draw[->] (c4.195) |- (c4\value.west);
	
	\foreach \value in {1}
	\draw[->] (c5.195) |- (c5\value.west);
	\end{tikzpicture}
}
	\medskip
	\caption{Tool modules and dependencies}
	\label{fig:tool-mods}
\end{figure}

In the sections that follow we explain each of the above properties and detail how the tool works for every case.
The algorithms implemented in CryptoSolve are supported by a set of base libraries for critical symbolic capabilities such
as term representation, term rewriting, unification, and more.
Currently the modules available in the tool and a simplified representation of their relation to each other is described
in Figure~\ref{fig:tool-mods}.

\paragraph{Outline}
In the remainder of the paper we cover the current state and capabilities of the tool
without focusing on the theory behind it, which can be found in \cite{LinLynch20,frocos2021,frocos2021LL,MeadowsESORICS21}.
Where necessary, we provide a brief theoretical background and indicate aspects on which the tool is based.
The rest of the paper is organized as follows.
We provide a discussion of related work in Section~\ref{sec:related}.
A brief review of necessary background material is covered in Section~\ref{sec:prelims}.
An overview of the security modules, their use, capabilities, and pointers
to the theory behind these methods are given in \cref{sec:moo-representation,sec:symbolic-security}.
The invertibility checking module is covered in Section~\ref{sec:invertibility}, the bounded authentication checking module is covered in Section~\ref{sec:authentication}.
We provide preliminary experimental results in Section~\ref{sec:experiments}.
Finally, the conclusions and future work are discussed in Section~\ref{sec:conclusions}.

%For the interested reader, the appendix includes an overview of the symbolic base modules, Section~\ref{sec:symb-lib}, and a brief introduction to the various interfaces for the tool, Section~\ref{sec:interface}.

\section{Related Work}\label{sec:related}
Publicly available tools for the generation and testing of security
property of cryptographic algorithms (e.g. \cite{DBLP:conf/ccs/BartheCGKLSB13, DBLP:conf/crypto/CarmerR16, 10.1145/2810103.2813636, DBLP:conf/csfw/MalozemoffKG14})
are the most closely related work to ours.
Perhaps the first of these is the work by Barthe et al. \cite{DBLP:conf/ccs/BartheCGKLSB13}.  This paper describes a tool, ZooCrypt, designed for the analysis of chosen plaintext and chosen ciphertext security public-key encryption schemes built from trapdoor permutations and hash functions.  A ZooCrypt analysis of a cryptosystem consists of two stages.  In the first stage a symbolic analysis tool is used to search for attacks on the cryptosystem.  If none are found, the analysis enters the second stage, in which an automated theorem prover  is used to search for a security proof in the computational model.

Later work looked at applying symbolic techniques and incorporates computational soundness results
to prove computational security.
For example,  Malozemoff et al. \cite{DBLP:conf/csfw/MalozemoffKG14} provide a symbolic algorithm whose successful termination implies adaptive chosen plaintext security
of cryptographic modes of operation using the message-wise schedule.
These results are extended by Hoang et al. in \cite{10.1145/2810103.2813636} to symbolic techniques for proving adaptive chosen ciphertext security of modes.
Both papers also include software that implements symbolic algorithms for generating  cryptosystems and proving their security.
Other work by Carmer et al.  \cite{DBLP:conf/crypto/CarmerR16} gives  a symbolic algorithm for deciding security of garbled circuit schemes,
and includes a tool, Linisynth, that generates such schemes and verifies their security using the algorithm.

All these tools have one thing in common: they only implement the algorithms described in the paper
they accompany, and thus are intended mainly as proofs of concept, not as general tools for the generation
and analysis of algorithms. The goal of CryptoSolve, however, is to serve as a tool for designing and
experimenting with multiple types of cryptosystems, security properties, and algorithms.
Thus, for example, it includes libraries for techniques that may prove useful in application to more
than one cryptosystem, such as unification, variant generation,  and the automatic generation of recursive functions.
It is also extensible, allowing more libraries and algorithms to be added as necessary,
and it includes an optional graphical user interface to make interactions with it easier. Currently, it
can be applied to  three different properties (static equivalence to random, invertibility, and authenticity, using five
different algorithms) of cryptographic modes of operation.

There is also a large amount of related research in formulating and proving indistinguishability properties for the symbolic analysis of cryptographic protocols.  These  properties are analogous to the computational indistinguishability  properties used in  cryptography.  The main differences are that symbolic indistinguishability does not always imply computational security (see, for example
Unruh \cite{DBLP:journals/iacr/Unruh10b}),  and 2) the symbolic algorithms are optimized for protocols, not crypto-algorithms, so applying them directly is not always advisable.   Even so, the approaches used in symbolic protocol analysis can be helpful.  For example an  undecidability result in Lin et al. \cite{frocos2021} is based on an undecidability result for cryptographic protocols  analysis due to K{\"{u}}sters and Truderung \cite{DBLP:conf/stacs/KustersT07}. 
To facilitate this interaction between symbolic
protocol analysis and symbolic cryptography, we use a formal model and specification language, due to
Baudet et al. \cite{baudet2005computationally}, that is based on the the concept of \emph{frames}
used by the applied pi calculus \cite{Abadi2001}, one of the most popular formal languages used by tools for the formal analysis
of cryptographic protocols.

\section{Preliminaries}\label{sec:prelims}
We first need to briefly review some background material both on MOOs and symbolic security, and also on the underlying term rewriting theory used in the tool. We begin with with term rewriting and related concepts. Please note, additional background material on equational theories, rewriting, and unification can be found here~\cite{Baader98}.
\subsection{Terms, Substitutions, and Equational Theories}
Given a first-order signature $\Sigma$, a countable set $N$ of variables bound by the symbol $\nu$ , and a countable set of variables $X$ (s.t. $X \cap N = \emptyset$), the set of terms constructed from $X$, $N$, and $\Sigma$, is denoted by $T(\Sigma, N \cup X)$. Note that since $N$ is a set of bound
variables we can often treat these as constants in the first-order theory and thus won't apply substitutions to these bound variables.
A substitution $\sigma$ is an endomorphism of $T(\Sigma,  N\cup X)$ with only finitely many variables not mapped to themselves, denoted by
$\sigma= \{ x_1 \mapsto t_1, \dots, x_m \mapsto t_m \}$. %Given two terms $s$ and $l$, determining if there exists a substitution $\sigma$ such that $l\sigma = s$, and computing $\sigma$ if it exists is called the \emph{matching problem}.
Application of a substitution $\sigma$ to a term $t$ is written $t\sigma$.
\ignore{
Given two substitutions $\theta$ and $\sigma$, the composition $\sigma \circ \theta$ is the substitution denoted here by $\theta\sigma$ and defined such that $x(\theta\sigma) = (x\theta)\sigma$ for any $x \in X$.
The domain of $\sigma$ is $Dom(\sigma) = \{ x \in X ~|~ x\sigma \neq x \}$. The range of $\sigma$ is $Ran(\sigma) = \{ x\sigma ~|~ x \in Dom(\sigma) \}$. When $\theta$ and $\sigma$ are two substitutions with disjoint domains and only ground terms in their ranges, then $\theta\sigma = \theta \cup \sigma$.
Given a substitution $\sigma$ and a finite set of variables $V \subseteq X$, the restriction of $\sigma$ to $V$ is the substitution denoted by $\sigma_{| V}$ such that $x\sigma_{| V} = x \sigma$ for any $x \in V$ and $x\sigma_{| V} = x$ for any $x \in X \backslash V$.}

% \subsection{Equational Theories}
Given a set $E$ of $\Sigma$-axioms (i.e., pairs of $\Sigma$-terms,
denoted by $l = r$), the \emph{equational theory} $=_E$ is the
congruence closure of $E$ under the law of substitutivity.
Since $\Sigma \cap N = \emptyset$, the
$\Sigma$-equalities in $E$ do not contain any bound variables in $N$.
An $E$-unification problem with bound variables in $N$ is a set of $\Sigma \cup N$-equations $P= \{ s_{1} =^{?} t_{1} , \dots , s_{m} =^{?} t_{m}  \} $.
 A solution to $P$, called an
\emph{E-unifier\/}, is a substitution $\sigma$ such that $s_i \sigma
=_E^{} t_i \sigma$ for all $1 \leq i \leq m$. \ignore{ A substitution $\sigma$
is \emph{more general modulo\/} $E$ than $\theta$ on a set of
variables $V$, denoted as $\sigma \leq_{E}^V \theta$, if there is a
substitution $\tau$ such that $x \sigma \tau =_{E} x \theta$ for all
$x \in V$.}

The primary equational theory implemented in the tool is the theory of xor,
denoted as $E_{xor}$.
This theory can be represented as a combination of a rewrite system,
$R_{\oplus}$, and an associative and commutative (AC) equational theory, $E_{\oplus}$.
$E_{xor} = R_{\oplus} \cup E_{\oplus}$:
$R_{\oplus} = \{x \oplus x \rightarrow 0, ~x \oplus 0 \rightarrow x\}$,
$E_{\oplus} = AC(\oplus)$, over the signature $\Sigma_{\oplus} = \{\oplus/2, f/1, 0/0\}$.
We will often denote this as the $MOO_{\oplus}$ algebra and modes of operations defined in this algebra
as $MOOs_{\oplus}$.

\ignore{
A rewrite rule $\ell \rightarrow r$ is applied to a term $t$ by
finding a subterm $s$ of $t$ and a match $\sigma$ of $l$ and $s$,
i.e., a unifier of $l$ and $s$ that leaves $s$ unchanged, and then
replacing $s$ with $r\sigma$.  We say that a term is in \emph{normal
  form} if no rewrite rule can be applied.  We note that any term in
the $E_{\oplus}$ theory is reducible via a finite set of rewrite rules
to a normal form term that is unique up to AC equivalence.
%moved from section 3
A variant of a term $t$ is all the distinct terms that can be obtained
by applying rules in a rewrite system.
A term is called normal if it cannot be reduced further
in the specified rewrite system.
%Repeated above-For any position $p$ in a term $t$ (including the root position $\epsilon$), $t(p)$ denotes the symbol at position $p$, $t|_p$ denotes the subterm of $t$ at position $p$, and $t[u]_p$ denotes the term $t$ in which $t|_p$ is replaced by $u$.
}

\subsection{Modes of Operation and Symbolic Security}\label{sec:symb-sec}
Before detailing the features of the tool we need to consider a few critical background notions such as Modes of Operation, Symbolic Security, and Authenticity. We do that in this section.
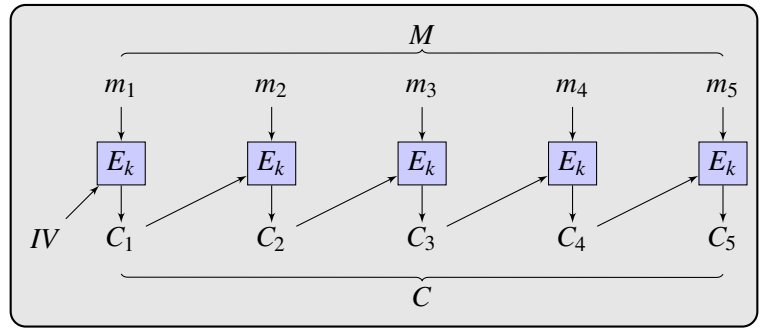
\begin{figure}
	\centering

	\tikzstyle{block} = [draw,fill=blue!20,minimum size=1.2em]
	\tikzstyle{surround} = [fill=black!10,thick,draw=black,rounded corners=2mm]
	\tikzstyle{brace} = [decorate,decoration={brace},thick]

	\begin{tikzpicture}[>=latex']

	\foreach \y in {1,2,3,4,5} {
		\node at (\y+\y,0) (input\y) {$m_\y$};
		\node[block] at (\y+\y, -1) (block\y) {$E_{k}$};
		\draw[->] (input\y) -- (block\y);
		\draw[->] (block\y.south) -- +(0,-0.5);
		\node at (\y+\y,-2) (cipher\y) {$C_\y$};
	}
	\foreach \x\y in {1/2,2/3,3/4,4/5} {
		\draw[->] (cipher\x) -- (block\y);
	}
	\node at (1, -2) (iv) {$IV$};
	\draw[->] (iv) -- (block1);

	\draw [decoration={brace,raise=5pt},decorate] (input1.north) -- node(M)[above=6pt] {$M$} (input5.north);
	\draw [decoration={brace, mirror, raise=5pt},decorate] (cipher1.south) -- node(C)[below=6pt] {$C$} (cipher5.south);

	\begin{pgfonlayer}{background}
	\node[surround] (background) [fit = (input1) (cipher5) (iv) (block5) (M) (C)] {};
	\end{pgfonlayer}

	\end{tikzpicture}
	\caption{An example block cipher to be modeled by a symbolic history}
	\label{Fig:Block-Cipher}
\end{figure}
\subsubsection{Modes and Their Security}
A cryptographic mode of operation can be described at a high level as follows.
The plaintext message $M$ is first broken into fixed sized blocks.
Each block $m_i$ is processed using the block encryption function $E_k$ along with some additional operations
to produce a  ciphertext block $C_i$. Typically,  the previous ciphertext is used in the computation of the current block,
and an initialization vector $IV$ is used to add randomness to the first block.
The final ciphertext is the sequence of ciphertexts thus produced.
Figure~\ref{Fig:Block-Cipher} illustrates this process for Cipher Block Chaining (CBC) mode.

%Many modes are possible, depending on how the mode of operation uses the randavailable prior components and  to produce the next ciphertext.
In order to model
these modes so that they can be checked via symbolic methods, we use
\emph{symbolic histories} (defined in~\cite{MeadowsESORICS21}). These describe interactions between
the adversary and the oracle, in which the adversary sends blocks of plaintext to be encrypted,
and the oracle sends back blocks of ciphertext according to
some fixed \emph{schedule} defined by the mode.
E.g., in a block-wise schedule a ciphertext block is sent immediately after it is generated by the mode. In a message-wise schedule,  the ciphertext blocks are not sent until  after the entire message is encrypted.

The symbolic definition of security we use is based on the computational security property {\sf IND\$-CPA} introduced by Rogaway in \cite{DBLP:conf/fse/Rogaway04}.
This is defined in terms of a game in which a challenger first chooses one of two oracles with probability 1/2.
The first is an encryption oracle that returns ciphertext when given plaintext,
and the second is a random bits oracle that returns a string of random bits that is as long as the ciphertext would have been.
The adversary interacts with the oracle  by sending it plaintexts and receiving the oracle's response.
At any time it can stop the game and guess which oracle it is interacting with.
Its \emph{advantage} is defined to be $\mid .5 - p \mid$, where $p$ is the probability that the adversary guesses correctly.
A mode is {\sf IND\$-CPA}-secure if its advantage is negligible in some security parameter $\eta$,
where a function $g$ is said to be negligible if for every polynomial $q$ there is an integer $\eta_q$ such that $g(\eta) < frac{1}{q(\eta)}$ for all $\eta > \eta_q$.
In the case of modes of operation, the security parameter is the maximum of the block size and the key size.
The motivation for a definition of this sort is that if the adversary cannot distinguish the output of the cryptosystem from random noise,
then it learns nothing about the plaintext.
This form of security, in which the security of a cryptosystem is quantified
in terms of the adversary's inability to distinguish between the output of an encryption oracle
and the output of an oracle that does not use the content of the plaintext in its calculations, is common in cryptography.

We note that if the adversary can create plaintexts that consistently cause a set of ciphertexts to exclusive-or to zero,
then it can distinguish between the real and random case with overwhelming probability.
If such an equality holds for the case in which the substitution is the identity,
we say that the mode is \emph{degenerate}.
In all other cases it is necessary but not sufficient that
the adversary must be able to consistently cause
at least one given pair of $f$-rooted terms to be equal,
known as a \emph{collision}.
We describe the symbolic model below, and then describe the unification problem that is associated with it.

\subsubsection{The Symbolic Model and Symbolic Security}
The blocks sent between the adversary and the oracle are modeled by terms in the \emph{$MOO_{\oplus}$ algebra}.
These \emph{$MOO_{\oplus}$-terms}  consist of free variables representing plaintext
blocks, bound variables representing a bitsting,
and terms built up using these variables and the signature $\Sigma=\{\oplus /2, 0/0, f/1 \}$,
under the Xor equational theory, where $f$ is the encryption function for some fixed key $K$, i.e., $enc(K, \_) = f(\_ )$.
Note that $f$ is not computable by the adversary.
%The terms of the frame are further restricted depending on the mode of operation  being modeled.
%The mode of operation dictates how the oracle constructs ciphertexts.
%For example, in  Cipher Block Chaining, $CBC$, the $i^\mathrm{th}$ ciphertext, $C_i$, is modeled by
%$f(C_{i-1} \oplus x_i)$, where $x_i$ is the
%$i^\mathrm{th}$ plaintext sent by the adversary.

A \emph{symbolic history} of the adversary's interaction with the oracle is modeled by a list of $MOO_{\oplus}$-terms
of the form $[t_1, ~t_2, \ldots, ~t_n]$.
All $MOO_{\oplus}$-terms are listed in the order that they are sent.
For example, the following symbolic history models the Cipher Block Chaining (CBC)
mode of operation with three ciphertexts using the block-wise schedule:
$\nu IV[IV, x_1, f(IV \oplus x_1), x_2, f(x_2 \oplus f(IV \oplus x_1))]$.
Here $IV$ is a bound variable representing an initialization vector. Each $x_i$ models a
plaintext block sent by the adversary and each $f$-rooted term is a ciphertext returned by the oracle according to the definition of the mode.
For example, in   $CBC$ the $i^\mathrm{th}$ ciphertext $C_i$ is modeled by
$f(C_{i-1} \oplus x_i)$, where $x_i$ is the
$i^\mathrm{th}$ plaintext.

Each symbolic history models the interleaving of one or more  \emph{sessions} between the adversary and oracle, where a session is a history that encrypts a single message consisting of a sequence  of plaintext blocks.
%The adversary has the ability to execute multiple simultaneous sessions with the oracle.
In this case the initial nonces, the $IVs$, will be fresh for each session.

The notions of  \emph{computable substitutions} and \emph{symbolic security}  are defined by Lin et al. in \cite{frocos2021}.
Let $P$ be a symbolic history. A substitution $\sigma$ is \emph{computable} w.r.t. $P$ if $\sigma$ maps
each variable $x_i$ to a term built up using the operators $0$ and $\oplus$ only on terms
returned by the oracle prior to receiving $x_i$ in $P$.
A mode of operation $M$ is \emph{symbolically secure} if there is no symbolic history $P$ of $M$
such that there is a non-empty set of terms $S$ returned by the encryptor %(called \emph{oracle terms})
in $P$ where $\bigoplus_{t \in S} t\sigma =_\oplus 0$,
and $\sigma$  is computable substitution w.r.t. $P$.
It is shown in \cite{frocos2021} that a mode of operation $M$
is symbolically secure if and only if $M$ is \emph{statically equivalent to random};
static equivalence \cite{Abadi2001} is a symbolic definition of indistinguishability
commonly used in symbolic protocol analysis.

We note that if a mode satisfies {\sf IND\$-CPA},
then it must be symbolically secure,
because if the adversary could make a substitution to the plaintext
such that it always satisfies the same equation by the ciphertext,
then it could easily distinguish the ciphertext from random with overwhelming probability.
A stronger condition has been shown by Meadows in \cite{MeadowsESORICS21} to imply {\sf IND\$-CPA}  security.
It has two parts. The first is non-degeneracy, which requires that symbolic security hold for the trivial case
in which the computable substitution $\sigma$ is the identity.
The second is the condition that no two different $f$-rooted terms have a computable unifier,
whether or not it leads to a violation of symbolic security.
This does not necessarily mean that symbolically secure modes that fail to satisfy the second condition
are not {\sf IND\$-CPA} secure, simply that more work may be required to prove them so.

\subsubsection{Checking Symbolic Security: Examples}\label{exp:symbolic-security}
Let's consider several examples of symbolic histories and checking for symbolic security. We start with the classic example of an insecure mode: the Electronic Code Book (ECB) mode. In ECB, each block is encrypted separately, so plaintext $x_1, \ldots, x_n$ yields ciphertext $f(x_1), \ldots, f(x_n)$.
\begin{figure}
	\centering
	\begin{subfigure}[b]{0.3\textwidth}
		\includegraphics[width=\textwidth]{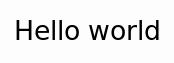}
		\caption{Image before ECB encryption}
		\label{fig:beforeecb}
	\end{subfigure}
	~ %add desired spacing between images, e. g. ~, \quad, \qquad, \hfill etc.
	%(or a blank line to force the subfigure onto a new line)
	\begin{subfigure}[b]{0.3\textwidth}
		\includegraphics[width=\textwidth]{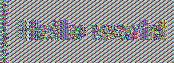}
		\caption{Image after ECB encryption}
		\label{fig:afterecb}
	\end{subfigure}
	\caption{ECB encryption with AES 128 ECB}\label{fig:ECB-Example}
\end{figure}
Notice that after applying ECB, the image in Figure \ref{fig:ECB-Example} is still not completely scrambled and some information from the original picture can still be deduced. This is because whenever two plaintext blocks are identical they produce the same ciphertext blocks.  Thus, any substitution unifying any two free variables is computable and  leads to a violation of symbolic security.

Other MOOs may be symbolically secure or insecure depending on the schedule. For example, consider a symbolic history of CBC with three ciphertext blocks: $P_2 = \nu IV[IV, x_1, f(x_1 \oplus IV), x_2, f(x_2 \oplus f(x_1 \oplus IV))]$. We consider two schedules:  the block-wise schedule, where each ciphertext block is returned to the adversary as soon as it can be computed, and the message-wise schedule, where they are returned  all together at the end. Note that in the block-wise schedule there is a computable unifier of $f(x_1 \oplus IV)$ and  $f(x_2 \oplus f(x_1 \oplus IV))$, namely $\sigma = \{ x_1 \mapsto IV, x_2 \mapsto f(0) \}$, but this is not computable in the message-wise schedule, which can be shown to be symbolically secure and {\sf IND\$-CPA}  secure.
%Rest of paragraph is ignored, but might be useful later in the paper
\ignore{It would then apply the $f$-rooted unification algorithm in an attempt to identify collision. Notice that for the block-wise schedule the CryptoSolve could identify a collision, $(f(x_1 \oplus IV) \oplus f(x_2 \oplus f(x_1 \oplus IV))\sigma =_{\oplus} 0$, where $\sigma = \{ x_1 \mapsto IV, x_2 \mapsto f(0) \}$. However, the tool needs to have seen the first ciphertext to identify the correct substitution to cause the collision. This would not be possible if all blocks were just returned in the end. Thus $CBC$ is symbolically insecure for the block-wise schedule. See~\cite{MeadowsESORICS21, LinLynch20, frocos2021} for a more detailed discussion of symbolic security.
Notice also that $\sigma$ is computable w.r.t. $P_2$ since $IV$ appears earlier than $x_1$ in $P_2$, and $f(x_1 \oplus IV)$ appears earlier than $x_2$ in $P_2$.}

Finally, we consider one additional MOO, Output Feedback Mode (OFB). OFB starts with an initialization vector, IV, each consecutive block, $C_{i+1}$, is computed as 
$C_{i+1} = T_{i+1} \oplus x_{i+1}$, where $x_i$ is the $i^{\text{th}}$ plaintext block, 
and $T_{i} = f(T_{i-1})$ ($T_0$ being IV).
For example, the first block would be $C_1 = f(IV) \oplus x_1$, and the second is
$C_2 = f(f(IV)) \oplus x_2$. Consider an OFB history with three ciphertext blocks: $P_3 = \nu IV[IV, x_1, f(IV) \oplus x_1, x_2, f(f(IV)) \oplus x_2]$. Note that, in order to unify   $f(IV) \oplus x_1$ and $f(f(IV)) \oplus x_2$, the adversary would have to set $\sigma x_2 = x_1 \oplus f(IV) \oplus f(f(IV))$, which it cannot do no matter what schedule is used, because it does not learn $f(f(IV))$ until after it has computed $x_2$. OFB is also both symbolically secure and {\sf IND\$-CPA}  secure. Notice that when generating the ciphertexts for differing MOOs such as CBC and OFB, the root symbol of the ciphertexts could differ and this will impact the unification algorithm required.
%Rest of paragraph is ignored, but might be useful later in the paper.
\ignore{The CBC oracle only returns bound variables and $f$-rooted terms, while the OFB oracle only returns bound variables and $\oplus$-rooted terms. In \cite{LinLynch20}, two unification problems are formalized: $f$-rooted local unification and $\oplus$-rooted local unification. $f$-rooted local unification problems arise when checking
security modes such as CBC, where all the oracle terms are either bound variables
or $f$-rooted terms. $\oplus$-rooted local unification problems arise when checking security of modes such as
of OFB, where all the oracle terms are either bound variables or $\oplus$-rooted terms.  We note that, since modes are defined recursively based on the signature $\Sigma$,
they generally fall into one camp or another and thus the tool selects the appropriate algorithm based on the MOO definition. See \cite{LinLynch20} for full details.}

\section{MOO Representation}\label{sec:moo-representation}
The tool contains a library implementation which allows for the representation and generation of $MOO_\oplus$-Programs.
The library currently allows $MOO_\oplus$-Programs that are constructed over the signature  $\Sigma=\{\oplus/2, 0/0, f/1\}$
and represented as a simple recursive function.
Once a $MOO_\oplus$-Program has been defined, the library can
then apply a number of operations on that $MOO_\oplus$-Program, including:
generating terms in a run of the $MOO_\oplus$-Program, checking symbolic security of the program, and checking invertibility.

\subsection{Standard and Custom $MOO_\oplus$-Programs}
Currently there are several well-known cryptosystems implemented to serve as examples for users when they are initially learning the tool. They also provide syntax examples for those wanting to add custom MOOs.
For example, the ciphertext chaining cryptosystem is defined below:

\begin{lstlisting}[title=Code]
from symcollab.moe import MOO
@MOO.register('cipher_block_chaining')
def cipher_block_chaining(iteration, nonces, P, C):
	f = Function("f", 1)
	i = iteration - 1
	if i == 0:
		return f(xor(P[0], nonces[0]))
	return f(xor(P[i], C[i-1]))
\end{lstlisting}

Notice that this provides a relatively simple example of the type of recursive cryptosystems built over an xor-theory that are currently supported. Here the base ciphertext is defined as $f(P_0 \oplus nonces(0))$, where
$P_0$ is the initial
plaintext sent by the adversary, and $nonces[0]$ is a bound variable representing the initialization vector. Then the recursive case is $C_i = f(P_i \oplus C_{i-1})$. The underlying libraries have been constructed to allow the encoded version of the system definition to closely match the theoretical one.%in such a way as

Similarly, a user can create their own custom mode of operation by adding the recursive definition to the MOO library.
\subsection{User defined schedule}
In addition to the block-wise and message-wise schedules (as described in Section \ref{sec:symb-sec}),
the user can define their own schedules based on the iteration number.
For example, this is a custom schedule that has the oracle only return ciphertexts on even iterations.
\begin{lstlisting}[title=Code]
from symcollab.moe import MOO_Schedule
@MOO_Schedule.register('even')
def even_schedule(iteration: int) -> bool:
	return iteration % 2 == 0
\end{lstlisting}

\subsection{Automatically Generated Singly Recursive $MOO_\oplus$-Definitions}\label{sec:moo-gen}
A user can ask the library to generate a recursive definition
of a modes of operation.
Currently there is one method in the tool library to automatically generate MOOs. It works by recursively generating
MOOs starting with the base components (IV, variables) and building singly recursive definitions using the xor and $f$ function, and recursive references to prior ciphertexts. The current method has some limitations. For example only one nonce is used, the signature is limited to $\Sigma = \{\oplus/2, 0/0, f/1\}$, only single recursion is used, and the base case is fixed to the initialization vector. Thus, the current method won't generate all possible $MOO_{\oplus}s$. For example, a MOO that uses two nonces in
its recursive definition won't be generated. We plan to expand this functionality in future versions of the tool allowing a user to automatically generate more classes of MOOs. Note, this doesn't limit the possible
MOOs that a user can analyze by using the custom module.
\ignore{
\noindent
\begin{minipage}[t]{.45\textwidth}
	\begin{lstlisting}[title=Code]
		from symcollab.moe import MOOGenerator
		gen = MOOGenerator()
		next(gen)
	\end{lstlisting}
\end{minipage}
\begin{minipage}[t]{.45\textwidth}
	\begin{lstlisting}[title=Output]
		f(P[i])
	\end{lstlisting}
\end{minipage}
}
The user can also filter the recursive definitions by properties such
as availability of the initialization vector, if it requires chaining,
or if the number of calls to the encryption function $f$ is less than
a specified bound. A mode of operation has the chaining property
if it incorporates a previous ciphertext into its recursive definition.

\ignore{
\noindent
\begin{minipage}[t]{.45\textwidth}
	\begin{lstlisting}[title=Code]
		from symcollab.moe import FilteredMOOGenerator
		gen = FilteredMOOGenerator(
		max_f_depth= 3,
		requires_iv=True,
		requires_chaining=False
		)
		next(gen)
	\end{lstlisting}
\end{minipage}
\begin{minipage}[t]{.45\textwidth}
	\begin{lstlisting}[title=Output]
		xor(IV, P[i])
	\end{lstlisting}
\end{minipage}
}

After creating the recursive $MOO_\oplus$ definition, we can then pass it to the class \texttt{CustomMOO}.
By default, this creates a $MOO_\oplus$ program with the specified recursive definition and
a new nonce for each base case.

\subsection{Interactions with $MOO_\oplus$-Programs}
Once a mode of operation and schedule have been defined, the tool can do several things with the definition. The first and simplest is to generate the terms corresponding to the symbolic representation of the ciphertexts. The user can also ask for the tool to evaluate the symbolic security of the MOO and/or the invertibility. We consider these options in the following sections. Before we move to security let's consider an example.

\paragraph{Examples}
The example MOOs in Table \ref{experiments:gen-not-all-secure} showcases ones that were generated
by the tool using the automatic generation feature.
Note that these are just a few examples.  In fact, one could allow the automatic generation to run as long as one wanted.

\begin{table}%\label{experiments:gen-not-all-secure}
	\begin{center}
		\small
		\begin{tabular}{ |l|l| }
			\hline
			\multicolumn{2}{|c|}{MOOs generated via Automatic Generation} \\
			\hline
			1 & $
			C_0 = IV,
			C_i = f(f(P[i]) \oplus P[i-1])
			$
			\\ \hline
			2 & $
			C_0 = IV,
			C_i  = f(f(P[i])) \oplus P[i-1] \oplus r
			$ \\ \hline
			3 & $
			C_0  = IV,
			C_i  = f(f(P[i]) \oplus C[i-1]) \oplus C[i-1]
			$ \\ \hline
			4 & $
			C_0 = IV,
			C_i = f(f(P[i]) \oplus C[i-1] ) \oplus f(f(P[i]) \oplus f(C[i-1]))
			$\\
			\hline

		\end{tabular}
	\end{center}
	\caption{Examples of MOOs generated by the automatic MOO generator}
	\label{experiments:gen-not-all-secure}

\end{table}

From these examples, the first two MOOs are not symbolically secure, they can be discovered and discarded by the method covered in the next section.
The final MOO is symbolically secure, however it is still useless since it doesn't have the invertibility property! This can also be checked via the method detailed below.
The third MOO, passes both the security and invertibility check and could be a candidate MOO for some secure application.

\section{Checking Symbolic Security Properties}\label{sec:symbolic-security}
This part of the tool is based on the work developed in~\cite{DBLP:journals/iacr/Meadows20, frocos2021}.
Those papers define a method for checking symbolic security which in turn can be used
to synthesize secure cryptographic modes of operation.
See Section~\ref{sec:symb-sec} for more background details.
We give an overview of each of the components developed for checking symbolic security beginning with
the $MOO_{\oplus}$-Programs.

\subsection{Checking Symbolic Security}
The tool can check for symbolic security in several ways. The first, and most exhaustive, is via the
local unification approach. In this approach ciphertexts of the $MOO_\oplus$-program under consideration are generated
and the appropriate local unification algorithm is used to see if any blocks sum to $0$, see~\cite{MeadowsESORICS21} for the full details of this approach.

The difficulty with this approach is that it can be time consuming in practice. However, a second approach has been developed in \cite{frocos2021}. The approach doesn't require
the generation of ciphertexts and works directly with the initial
$MOO_\oplus$-program definition. This approach is not complete, but it works for many cases and has the advantage of being much more efficient. Therefore, we have implemented it as a first pass symbolic security
check for the tool. If the first pass cannot decide symbolic security, then, the full security check
requiring block generation will be used.

\paragraph{Examples}
Continuing With the MOOs from Table~\ref{experiments:gen-not-all-secure}, let's consider just the first MOO.
We can check for security using the MOO check method:
\begin{lstlisting}[title=Code]
moo_check(moo_name = 'table1.1', schedule_name = 'every',
	unif_algo = p_syntactic, length_bound = 10, knows_iv = True,
	invert_check=True)
\end{lstlisting}

The tool would return the first collision it finds, which violates the symbolic security property, for this example:
\begin{lstlisting}[title=Output]
Here is the problem:
f(xor(f(x1), IV)) = f(xor(f(x2), x2))
\end{lstlisting}

%\todo[inline]{Output does not show collision}

\section{Invertibility and Recovering the Plaintext}\label{sec:invertibility}
A cryptographic algorithm is \emph{invertible} if given a ciphertext and a decryption
key, the original plaintext can be retrieved. This is not a given for any $MOO_\oplus$-program, even a secure one. Therefore, in the automatic generated setting we will need
methods for checking if the invertibility property holds for any particular $MOO_\oplus$-program.
Currently the tool is able to check invertibility for a large class of recursively defined MOOs. This class includes the well known MOOs, such as CBC, ECB, and CFB. More detailed information on theory and method for checking invertibility has been presented in~\cite{frocos2021}.

The invertibility checker is built into the MOO security check functionality in the tool and can be requested simply by setting the ``invert\_check'' flag (which is the last flag) in the $moo\_check$ function. See Example~\ref{example:invert}.

\begin{example}\label{example:invert}\end{example}%\vspace{-.5cm}
\begin{lstlisting}[title=Code]
from symcollab.moe.check import moo_check
from symcollab.Unification.p_unif import p_unif
result = moo_check('cipher_block_chaining', "every", p_unif, 2, True, True)
print(result.invert_result) # prints True
\end{lstlisting}

\section{Authentication} \label{sec:authentication}
An authenticated encryption scheme \cite{xcbc, ocb} satisfies the authenticity property if an adversary cannot forge any new valid ciphertext message after observing any number of valid ciphertext messages. In \cite{frocos2021LL}, the authors proposed two algorithms for checking authenticity. The first algorithm works for a simplified case, where only messages of fixed length can be handled. The second algorithm works for the general case, where messages of arbitrary length can be handled.

We use $M_1, M_2, \ldots\ldots$ to denote plaintext blocks, and use $C_1, C_2, \ldots\ldots$ to denote ciphertext blocks.
$\mathcal{E}_K(T, \cdot)$ denotes a tweakable block cipher \cite{tweakableBC}, where $K$ is some key and $T$ is some tweak. The idea is that each key and tweak produce an independent pseudorandom permutation. $\mathcal{D}_K(T, \cdot)$ is the inverse of $\mathcal{E}_K(T, \cdot)$. $n(T)$ produces another tweak, given a tweak $T$.
We use $n^k(T)$ as a shorthand for applying $n$ to $T$ for $k$ times. The idea is that the same key can be used for multiple blocks, as long as the tweaks are different for each different block. In order to achieve authenticity, a tag is attached to each ciphertext message.
Each scheme is associated with a \textit{verification condition}, which refers to the ciphertext blocks and the tag.
A ciphertext message is \textit{valid}, if the verification condition holds for that ciphertext message.

\begin{figure}
\centering
\begin{subfigure}{.5\textwidth}
  \centering
  \includegraphics[width=.8\linewidth]{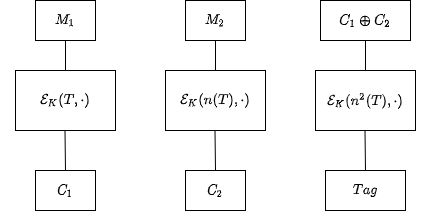}
  \caption{An Insecure Scheme}
  \label{fig:insecure-ae}
\end{subfigure}%
\begin{subfigure}{.5\textwidth}
  \centering
  \includegraphics[width=.8\linewidth]{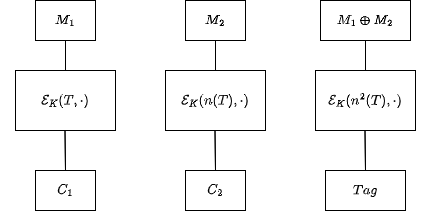}
  \caption{A Secure Scheme}
  \label{fig:secure-ae}
\end{subfigure}
\caption{Two Authenticated Encryption Schemes}
\label{fig:ae-schemes}
\end{figure}

Figure \ref{fig:ae-schemes} shows two authenticated encryption schemes, both of which handle messages of two blocks.
The verification condition of the scheme in Figure \ref{fig:insecure-ae} is $\mathcal{E}_K(n^2(T), C_1 \oplus C_2) = Tag$. The authenticity property is violated. The reason is that if $(C_1, C_2, Tag)$ is a valid ciphertext message, $(C_1 \oplus C_2, 0, Tag)$ is also a valid ciphertext message. The verification condition of the scheme in Figure \ref{fig:secure-ae} is $\mathcal{E}_K(n^2(T), \mathcal{D}_K(T, C_1) \oplus \mathcal{D}_K(n(T),C_2)) = Tag$, the authenticity property is satisfied. The intuition is that the adversary does not know any way of modifying $C_1$ and $C_2$ in such a way that $M_1 \oplus M_2$ remains the same.

Here are some other possible verification conditions for authenticated encryption schemes, which handle two message blocks. The first three verification conditions satisfy the authenticity property, and the last two do not.

\begin{itemize}
\item $\E_K(n^2(T), \D_K(n(T), \D_K(T, C_1) \oplus C_2)) = Tag$
\item $\E_K(n^2(T), \D_K(T, \D_K(n(T), C_2) \oplus C_1)) = Tag$
\item $\E_K(n^2(T), \D_K(T, C_1) \oplus \D_K(n(T), C_2)) = Tag$
\item $\E_K(n^2(T), \D_K(T, C_1)) = Tag$
\item $\E_K(n^2(T), \D_K(n(T), C_2)) = Tag$
\end{itemize}

%As the number of message blocks gets bigger, there are more schemes and verification conditions to check. 
Given a verification condition of some authenticated encryption scheme, our tool can automatically check if the authenticity property is satisfied.

\noindent
\begin{minipage}[t]{.70\textwidth}
\begin{lstlisting}[title=Code]
from symcollab.Unification.constrained.authenticity import *
t = e(n(n(T)), xor(C1, d(n(T), C2)))
check_security(t)
\end{lstlisting}
\end{minipage}
\begin{minipage}[t]{.25\textwidth}
\begin{lstlisting}[title=Output]
The authenticity property is satisfied.
\end{lstlisting}
\end{minipage}

\section{Experiments}~\label{sec:experiments}
A benefit of the tool design is that it is easy to integrate the above described functions into a
script which can then be used to run experiments.
For example, we have included a script, located in the experiments directory of the tool,
that allows the user to run longer experiments and can handle restarts.
In this script, we generate new candidate MOOs one at a time and
test them for security. The output of \texttt{moo\_check} is the data structure called \texttt{MOOCheckResult}.
This has the following fields: \texttt{collisions} (set of computable substitutions that cause a collision to occur),
\texttt{invert\_result} (whether or not the MOO is invertible), \texttt{iterations\_needed} (number of iterations before a collision was found),
and whether or not the MOO satisfies symbolic security up to the bound checked.

\paragraph{Initial Experimental Results}\label{sec:results}
A sample of some of the secure MOOs found during early experiments are listed in Table \ref{table:experiments-gen}. All of these MOOs were created automatically by the currently implemented recursive \texttt{MOOGenerator}.
As a future work, we plan to create additional generators that the user can select and allow for user defined generators.

\begin{table}\label{experiments:gen}
	\begin{center}
		\small
		\begin{tabular}{ |l|l| }
			\hline
			\multicolumn{2}{|c|}{Secure MOOs Found via Automatic Generation and Testing} \\
			\hline
			1 & $
			C_0 = IV,
			C_i = f(f(f(P[i-1]) \oplus r) \oplus C[i-1])
			$
			\\ \hline
			2 & $
			C_0 = IV,
			C_i  = f(f(f(P[i])) \oplus C[i-1] \oplus r)
			$ \\ \hline
			3 & $
			C_0  = IV,
			C_i  = f(f(P[i]) \oplus C[i-1]) \oplus C[i-1]
			$ \\ \hline
			4 & $
			C_0 = IV,
			C_i = f(f(f(P[i]) \oplus r \oplus C[i-1]))
			$ \\ \hline
			5 & $
			C_0 = IV,
			C_i = f(f(P[i]) \oplus C[i-1]) \oplus f(C[i-1])
			$\\
			\hline

		\end{tabular}
	\end{center}
	\caption{Examples of secure MOOs found using the MOO generator}
	\label{table:experiments-gen}
\end{table}

Experiments can also be done without the \texttt{MOOGenerator}, where MOOs are generated via hand or a custom script and then checked for security. This is an attractive option because it allows the user to easily customize the type of MOOs they are considering. Table~\ref{table:experiments-custom} includes some example secure MOOs that were created by hand and then tested for security using the tool. Note, that although all three MOOs are secure only the first MOO can be shown by the tool to be invertible (via the method developed in~\cite{frocos2021})! Thus, secure but useless MOOs can also be discarded.

\begin{table}
	\begin{center}
		\small
		\begin{tabular}{ |l|l| }
			\hline
			\multicolumn{2}{|c|}{Secure MOOs Found via Custom Generation and Testing} \\
			\hline
			1 & $
			C_0 = IV,
			C_i = f(P[i] \oplus f(C[i-1])) \oplus f(C[i-1])
			$
			\\ \hline
			2 & $
			C_0 = IV,
			C_i  = f(P[i] \oplus f(C[i-1]) \oplus f(P[i]))
			$ \\ \hline
			3 & $
			C_0 = IV,
			C_i  = f(f(P[i]) \oplus C[i-1] ) \oplus f(f(P[i]) \oplus f(C[i-1]))
			$ \\ \hline

		\end{tabular}
	\end{center}
	\caption{Examples of secure MOOs found using a custom generator}
	\label{table:experiments-custom}
\end{table}

Based on the initial experimentation with the tool there are some interesting early questions:
Can the set of secure MOOs be closed under some operation such as applying the encryption symbol $f$ on top? Are there cases where we can place a bound on the number of iterations to check security?
We're particularly motivated by the second question, due to the complexity
of our saturation based decision procedures. For some of the MOOs we tested,
it took in the order of days in order for the algorithm to find a collision.

\section{Conclusion and Future Work}\label{sec:conclusions}

In this paper we presented a new tool for the symbolic analysis of cryptosystems
with the ultimate goal of providing support for multiple types of algorithms and
representations.
Although at present it only supports modes of operation,
the tool provides a widely applicable symbolic foundation based on that of Baudet et al. \cite{baudet2005computationally}.
Not only can this symbolic foundation represent multiple types of cryptosystems, the symbolic foundation is also
amenable to proofs of computational soundness and completeness.
The tool also includes libraries that support useful algorithms for checking symbolic security, including unification and variant generation.
There are limitations to the currently supported modes of operation. Currently, only modes which can be modeled using the above Xor theory are supported with the needed 
specialized unification algorithms. For example, modes requiring primitives such as hash functions, successor, or full abelian groups are currently not supported. However, we
hope to add, if possible, support for as many of these structures as possible in future 
versions of the tool.

One avenue of interest to us is to investigate other work on symbolic cryptography
to determine whether it can be fit into our framework and its algorithms implemented in our tool.
We expect previous work on modes \cite{DBLP:conf/csfw/MalozemoffKG14,10.1145/2810103.2813636} to fit in well,
since, although their models are not expressed as symbolic algebras, they are still compatible with the algebra used in CryptoSolve.
Other work, such as Linicrypt \cite{DBLP:conf/crypto/CarmerR16,DBLP:conf/tcc/McQuoidSR19} and Zoocrypt \cite{DBLP:conf/ccs/BartheCGKLSB13}
also follow the paradigm of representing cryptographic primitives as function symbols obeying equational theories.
In the cases in which soundness and completeness results are provided, we expect them to carry over into the symbolic model.
When computational soundness of a symbolic language is not known (e.g. Zoocrypt),
it may be possible to ensure it by limiting its expressiveness.

More generally, our tool is intended to be extensible to cryptosystems that may not yet have been studied from the symbolic point of view.
Although only a few function symbols have been implemented in CryptoSolve as of now,
it is designed to be extensible.
The best choice, for this seems to be cryptosystems that can be expressed in terms of combinations of primitives, including randomly chosen bitstrings, each of which has a clearly defined security property.
The combinators should be operations that can characterized in a symbolic way.
These include not only finite field and group operations, which are commonly used in cryptography,
and can be found in all the work cited in this paper, but concatenation,
which is used in \cite{DBLP:conf/ccs/BartheCGKLSB13,DBLP:conf/tcc/McQuoidSR19,Hollenberg22}.
Many cryptosystems are defined using these techniques of building complex systems from basic components,
so we expect the are of application to be wide.

\bibliographystyle{eptcs}
\bibliography{MOE-bib}

\end{document}